\documentstyle[aps,prl]{revtex}
 
\begin{document}

\twocolumn[\hsize\textwidth\columnwidth\hsize\csname
@twocolumnfalse\endcsname
 
\title{Textures and Newtonian Gravity}

\author{ Eduardo Gu\'eron$^*$ and P.S. Letelier$^\dagger$} 
 
\address{
 Departamento de Matem\'atica Aplicada-IMECC,
Universidade Estadual de Campinas,
13081-970 Campinas,  S.P., Brazil}

\maketitle
 
\begin{abstract}
Newtonian theory is used to study the gravitational effects of a
 texture, in particular the formation of massive structures.
\vspace{0.4cm}

\end{abstract}


\vskip2pc]
 
\footnote[0]{$^*$Electronic address: gueron@ime.unicamp.br}
\footnote[0]{$^\dagger$Electronic address: letelier@ime.unicamp.br}

 Topological defects: domain walls  , strings and monopoles have been 
recently the focus of much attention in cosmology \cite{peebles}--\cite{brand}. 
The symmetry 
breaking of non-Abelian groups always leads to the formation
 of textures, a cosmic
defect that can be asymptotically represented as a space-time point, an
event \cite{davis,turok}. In this case the homotopy theory tells us that the
vacuum manifold topology is not trivial, $\pi _3(M)\approx Z$.

Textures are global defects that can give rise to seeds of cosmic
 background anisotropies \cite{ruth,micro}. In principle, we may observe 
its effect by looking to the
microwave spectra fluctuations, like the ones obtained by the satellite
 COBE \cite{cobe}. The aim of this note is to study the mass accretion
 by a texture in the simplest way. We shall see that the use of Newtonian 
gravity, the zeroth order approximation to general relativity, makes
possible to visualize the formation of massive structures in a
 very simple way.

The Lagrangian density associated to textures with $SU(2)$
 symmetry can be written as,

\begin{equation}
{\cal L}=\frac 12\partial ^\mu \Phi ^{\dagger a}\partial _\mu \Phi ^{
a}-\frac \lambda 4\left( \Phi ^{a}\Phi ^{\dagger a}-\eta ^2\right)
^2.  \label{lag}
\end{equation}
 
 Since, we are interested 
only in the scalar sector of the theory,   instead of SU(2) 
we shall  require SO(4) symmetry. Then, the scalar fields may be represented 
as a real quadruplet.
To break the symmetry  
we  introduce  in the action the constrain $\Phi ^a\Phi ^a=\eta ^2$ via
 a Lagrangian multiplier $\alpha $. Hence,  the symmetry will be
  broken from
$SO(4)$ to $SO(3)$.  This is possible because for textures 
the third homotopy group of the vacuum manifold must be non trivial  and $\pi_3(SO(4)/SO(3))\approx\pi_3(SU(2)/1)\approx Z$. 

The action in Minkowski spacetime  ($M_4$), 
\begin{eqnarray}
&&S=\int \frac 12\partial ^\mu \Phi ^a\partial _\mu \Phi ^a-\frac \lambda
4\left( \Phi ^a\Phi ^a-\eta ^2\right) ^2 \nonumber\\
&&\hspace{2cm}+\alpha (\Phi ^a\Phi ^a-\eta ^2)  d^4 x,
\label{action}
\end{eqnarray}
yields the texture field equations, 
\begin{equation}
\Box\Phi^a=-\frac{\nabla_\mu\Phi^b\nabla^\mu\Phi^b}{\eta ^2}\Phi^a .
\label{eqfield}
\end{equation}

 The constrain $\Phi ^a\Phi ^a=\eta ^2$ tells us that the  texture can be 
regarded  as a 3-sphere located in a four dimensional field 
space. Therefore, the most  natural parametrization \cite{davis} of $\Phi^a$ 
is 
\begin{eqnarray}
 &&\Phi^{(0)} =\eta \cos \chi (r,x^0)\nonumber\\
 &&\Phi^{(i)}=\eta[\sin \chi (r,x^0)\cos \varphi 
\sin \theta 
\sin \chi (r,x^0)\sin \varphi \sin \theta , \nonumber\\
&&\hspace{2cm}\sin \chi (r,x^0)\cos \theta ]
, \label{param}
\end{eqnarray}
where, $x^0=ct$. The motion equations  reduces 
to a single one  for the function $\chi(r,x^0)$,
\begin{equation}
\frac{\partial^2\chi}{\partial {x^{\tiny 0}}^{2} }-\frac{ 2}{r} 
\frac{\partial\chi}{\partial r}  - \frac{\partial^2\chi}{\partial r^2} =
\frac{\sin 2\chi }{r^2}\,.  \label{equ}
\end{equation}
The study of the  Lie symmetries of this equations  gives us   
the similarity variables, $u=r/x^0$, and $r$. The first one 
 reduces (\ref{equ}) to
\begin{equation}
\chi'' +\frac{ 2}{u}\chi' =
\frac{\sin 2\chi }{u^2(1-u^2)}\,.  \label{equ2}
\end{equation}
The prime indicates derivation with respect to the similarity 
variable $u=r/x^0$. 


It is a simple exercise to prove that the 
 geodesic equations for particles moving  with
non-relativistic speed ($v\ll c$) in the presence of  a  
weak gravitational field
$g_{\mu\nu}(x^\lambda)=\eta_{\mu\nu}+h_{\mu\nu}(x^{\lambda})$, 
  ($|h_{\mu\nu}|^2\ll 1$),
can be cast in the   the Newtonian form 
\begin{equation}
\frac{d^2{\bf{r}}}{dt^2}=-\frac{1}{2}c^2\nabla h_{00} .
\label{geod}
\end{equation}
Thus, in this approximation   the component $h_{00}$ plays, essentially, 
 the role of the Newtonian potential, and the other components play no role at all.
Therefore, we shall consider that the weak gravitational field
  associated with the texture can be 
described by the spherically symmetric metric,
 \begin{equation}
ds^2=[1+   h_{00}(t,r)]c^2dt^2-dr^2-r^2(d\vartheta^2+
\sin^2\vartheta d\varphi^2).  \label{metric}
\end{equation}
  
By solving the eigenvalue equation for the Ricci
tensor [$\det(R^\mu_\nu-\lambda\delta^\mu _\nu)=0$]
derived from  (\ref{metric}), we get for the eigenvalue
 associated to  the timelike eigenvector,
\begin{equation}
\lambda_0^R=(h_{00,rr}+2h_{00,r}/r)/2,  \label{ptrG}
\end{equation}
 that in this case is just $R^t_t$.

Recalling that the  eigenvalues values of matrices  $N$ and $M$
such that $M=N+\alpha I$ are related by $\lambda_M=\lambda_N +\alpha$, we 
have that the Einstein equations $R^\mu_\nu=\frac{8\pi
 G}{c^2}(T^\mu_\nu -\frac{1}{2}T \delta^\mu_\nu)$ give us for the 
eigenvalue related to the timelike eigenvector
\begin{equation}
\lambda^R_0=\frac{8\pi G}{c^2}(\lambda^T_0 -\frac{1}{2}T). \label{lambdas}
\end{equation}
The trace of the energy-momentum tensor (EMT) can be always written as 
the sum of its eigenvalues. The eigenvalue related to the 
timelike eigenvalue is defined as the material density times the square
 of the light velocity  $\rho c^2$, and the eigenvalues related to the
 spacelike eigenvectors are denoted as:  $-p_1, -p_2 $, and  -$p_3 $
 \cite{synge}. The
 quantities  $p_i$ when positive (negative) are the principal 
pressures (tensions). From (\ref{lambdas}),  (\ref{ptrG}) and the
 definition $\psi\equiv c^2 h_{00}/2$ we find
\begin{equation}
\nabla^2\psi=4\pi G\rho_N \label{lap}
\end{equation}
where $\rho_N$ is the associated Newtonian density,
\begin{equation}
\rho_N=\rho +(p_1+p_2+p_3)/c^2 . \label{ron}
\end{equation}
From (\ref{geod}) we find
  \begin{equation}
\frac{d^2{\bf{r}}}{dt^2}=-\nabla\psi. \label{motion}
\end{equation}

The concept of associated  Newtonian density of  a given distribution
 of matter was first
developed by Tolman \cite{tolman} for quasi-static metrics,  lately 
 MacCrea and Milne \cite{McM}  found the Friedmann-Lema\^\i the equation
 for the expansion
 rate of the universe radius using  Newtonian considerations  \cite{pee2}. 
For applications of this concept to cosmic strings and domain walls see 
Refs. \cite{vil}. 

We  shall begin our considerations about structure formation by 
studying the texture energy-momentum tensor. The metric 
 and the canonical EMT are equal
 in this case.  This tensor can be computed for the generic
solution of equation  (\ref{equ2}),  $\chi(r/t)$ (from now on we shall
use units such that $c=1$). We find, 
\begin{eqnarray}
 && \left( {\hat T}_\nu ^\mu \right) =\frac{ \eta ^2(\chi ^{\prime })^2}{ 2t^2}
 \times \nonumber \\ &&\left( 
\begin{array}{cccc}
1+u^2+F^2 & 2u^2 & 0 & 0 \\ 
-2u^2 &  -1-u^2+F^2 & 0 & 0 \\ 
0 & 0 & 1-u^2 & 0 \\ 
0 & 0 & 0 & 1-u^2 \\ 
  \end{array}
\right),  \label{emt1}
\end{eqnarray}
 where $F= 2 \sin ^2\chi/ (u\chi ^{\prime })^2.$  By
 solving the eigenvalue problem associated with (\ref{emt1}) we 
find its  diagonal form,

 \begin{eqnarray}
&&  \left(  T_\nu ^\mu \right) =\frac{ \eta ^2(\chi ^{\prime })^2}{ 2t^2}
 \times \nonumber \\ &&\left( 
\begin{array}{cccc}
|1-u^2|+F^2 & 0 & 0 & 0 \\ 
0 &  -|1-u^2|+F^2 & 0 & 0 \\ 
0 & 0 & 1-u^2 & 0 \\ 
0 & 0 & 0 & 1-u^2 \\ 
& & &
 \end{array}
\right). \label{tensor}
 \end{eqnarray}
 
  \vspace{0.2cm} 

\noindent 
The  EMT (\ref{tensor}) indicates that the associated gravitational
 density $\rho_N$ vanishes when $r<|t|$ and it is positive when $r>|t|.$
  The relation (\ref{ron}) tells  us that not only the mass density
 gravitates, but also the associated pressures contribute positively
 to this mass density and the tensions negatively. 
 This simple observation allows us to foresee some aspects of
 the microwave distortions produced by a texture. 
 
Let us consider  a texture  localized at the origin of $M_4$, $r=t=0$
and its corresponding light cone (Fig.1).

\begin{figure}[h]
\vspace{8cm}
\includegraphics{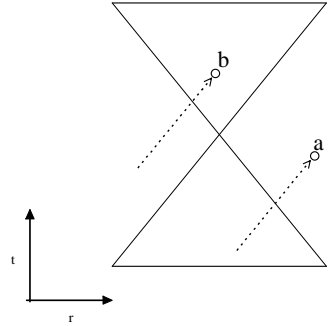}
\vspace{-1.6cm}
 \caption{ The photon $a$ is red-shifted by the presence of 
the texture at $t=r=0$ whilst the photon $b$ is blue-shifted. 
}
\end{figure}

  Photons entering  the  cone   are   attracted  towards the center, 
 their energy increase (they are blue-shifted). The  photons leaving the
 cone are also attracted,  but their  energy decrease
 (they are red-shifted). 
This 
result was obtained by Turok and Spergel \cite{micro} using the
 Einstein equations in the weak field
approximation  with the EMT associated to the special 
solution   of (\ref{equ2}),

\begin{equation}
\chi (r/t)=\left\{ 
\begin{array}{ll}
2\arctan (-r/t), & t<0 \\ 
2\arctan (r/t)+\pi & 0<r<t \\ 
2\arctan (t/r)+\pi & 0<t<r
\end{array}
\right. ,  \label{sol}
\end{equation}
which represents an unwinding texture that changes its topological 
charge at $t=0.$
 
Now we shall study the effect of the texture on matter using  Newtonian
gravitation, that is the zeroth order approximation to general relativity
Eqs. (\ref{lap})-(\ref{motion}). 
 By numerically solving the Barriola and Vaschaspati \cite{vasch} 
equations for a self-gravitating texture we find that the weak field
 condition is  fulfilled   \cite{thesis}. 
 
From (\ref{sol}), (\ref{ron}), and  (\ref{tensor}) we find  
\begin{equation}
\rho_N=\left\{ 
\begin{array}{ll}
0, & r\leq |t| \\ 
\frac{\textstyle 8\eta ^2(r^2-t^2)}{\textstyle (r^2+t^2)^2}, & r>|t|
\end{array}
\right. .  \label{dens}
\end{equation}
It is important to emphasize that the singularity for this solution
 is exactly at the point $(t=0,r=0)$. By solving (\ref{lap})
 with (\ref{dens}) as a source  we get  the
Newtonian gravity of the texture $\bf{G}=-\nabla\psi$, 
\begin{eqnarray}
 &&G_r =8\pi G\eta ^2\left[ \frac{[2r-3t +\pi t  -4t\arctan (r/t)}{2r^2} \right.
\nonumber\\
 &&\hspace{3cm}\left. +\frac{t^2}{r(r^2+t^2)}\right], \label{gr}
\end{eqnarray}
in the region $|t|<r$ and zero elsewhere.
 We can use (\ref{gr}) to make simple simulations to
 describe the gravitational action of a texture in the following way:
Firstly  we  numerically solve  the Newtonian motion equation for a
 particle subject to the central force per unit of mass $G_r$. The
 initial conditions are the ones for an homogeneous medium.
 When the time variable $t$  reaches the value of the  $i^{th}$
 particle radial coordinate,  $r_i$, we stop the
 integration.  At $t=r_i$ the density $\rho_N$ and the  associated
 gravitational potential are zero. Then, at this instant the
  particle starts to move with  constant velocity. This velocity
is taken to be the same computed previously. In Fig. 2 we
 show the initial position of a distribution  of particles. 
Fig. 3 presents a  mass concentration due to the attractive
 character of the gravitational force
of the texture.

\begin{figure}[hc]
\vspace{6cm}
\includegraphics{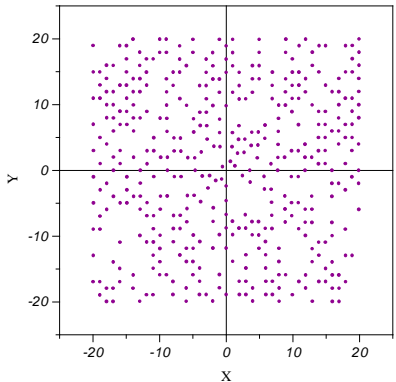}
\vspace{0.8cm}
 \caption{ Initial configuration.}
\end{figure}

\vspace*{7cm}

\begin{figure}[hc]
\includegraphics{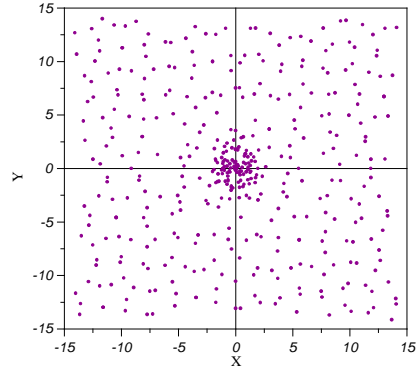}
\vspace{0.8cm}
 \caption{   Maximum density configuration.}
\end{figure}

 In Fig. 4 we  see the concentration
 of particles being diluted for $t$ larger than all $t_i$.
The  matter is less concentrated than the homogeneous initial form.

\begin{figure}[hc]
\vspace*{6cm}
\includegraphics{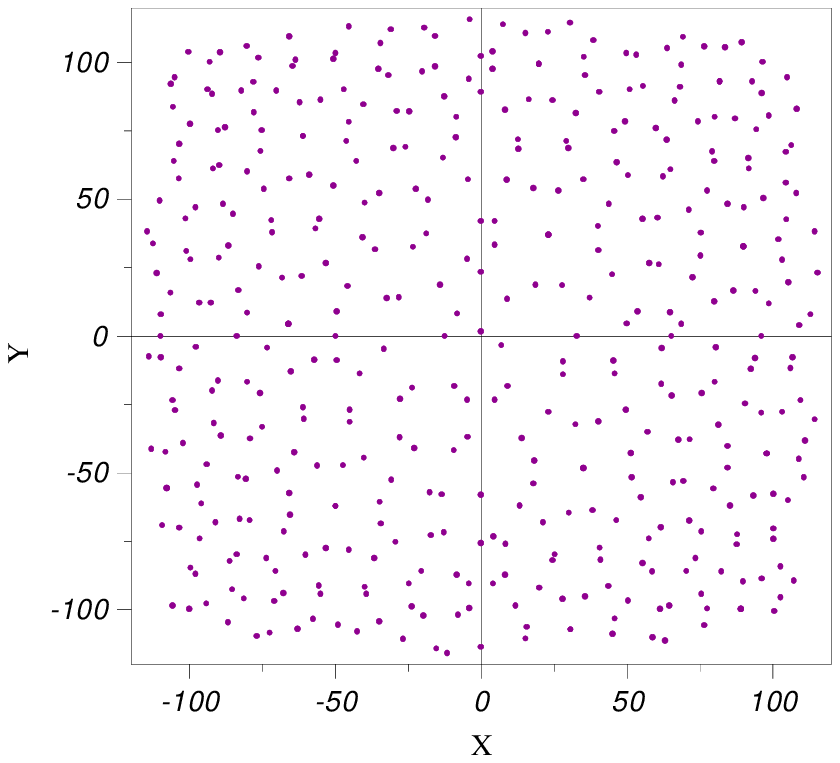}
\vspace{0.8cm}
 \caption{ Mass distribution after some time. This configuration expands 
forever.
 }
\end{figure}

 In the simulations the texture parameter $\eta$ has been greatly  
magnified, we  used  $\eta=10^{19}GeV$, a thousand times larger
 than the usual one.  The configurations shown in the graphics do
 not depend on the space time units. The time variation from Fig. 2 
to Fig. 4 is  approximately $10^{-6}$ seconds for the length unit meter.
 Also, in order to have 
a meaningful Newtonian picture, 
in the simulations the velocity of the test particles at most 
reached a few percents of the speed of light.  
Some authors simulate the radiation density fluctuation produced by a
 texture  using an  approximate  solution for small
 $u=r/t$ \cite{not,barr}.  This region is
exactly where the Newtonian density does not vanish. Therefore
 we can use our result to  estimate the matter density anisotropy
 produced by a  texture calculating the mass concentration at the
 instant represented in the Fig.3 and compare it  with the initial
 one, Fig. 2.

The $\eta $ value is the order of $10^{16}GeV $
 \cite{ruth}, using 
the Planck mass we get
  
\begin{equation}
\frac{\delta \rho _m}{\rho _m}\simeq 10^{-3}.  \label{est}
\end{equation}

Phillips \cite{phil} found that this kind of texture is not relevant  to
seed  the microwave background density fluctuation because
this event is much rarer than the 
less energetic ones. This fact might reduce the importance of this approach.
 
 However the main aspect to be considered is that events like the
Turok's texture are more energetic than  others because of the potential
 barrier of the unwinding process \cite{borril}. We can speculate 
 that the texture analyzed here can form an isolated 
 very massive structure, e.g. The
Great-Attractor \cite{lynd}. 

In summary using a very simple Newtonian simulation one can
 obtain a good qualitative picture of the structure 
formation due  to textures.

The authors thank  FAPESP,  CNPq and CAPES for financial support.

\end{document}